\documentclass{article}
\usepackage{spconf,amsmath,graphicx}
\usepackage{multirow}
\usepackage{makecell}
\usepackage{booktabs}
\usepackage{pifont}
\title{The NUS-HLT System for ICASSP2024 ICMC-ASR Grand Challenge}
\name{
\parbox{\linewidth}{\centering
Meng Ge$^{1}$, Yizhou Peng$^{1}$, Yidi Jiang$^{1}$, Jingru Lin$^{1}$, Junyi Ao$^{2,3}$, Mehmet Sinan Y{\i}ld{\i}r{\i}m$^{1}$, \\Shuai Wang$^{2,3}$, Haizhou Li$^{1,2,3}$, Mengling Feng$^{1}$\thanks{The computational work for this artical was partially performed on resources of the National Supercomputing Centre, Singapore (https://www.nscc.sg).}}}
\address{$^{1}$National University of Singapore, $^{2}$Shenzhen Research Institute of Big Data, Shenzhen, China \\ $^{3}$The Chinese University of Hong Kong, Shenzhen (CUHK-Shenzhen), China}

\begin{document}
\ninept
\maketitle
\begin{abstract}
This paper summarizes our team's efforts in both tracks of the ICMC-ASR Challenge for in-car multi-channel automatic speech recognition. Our submitted systems for ICMC-ASR Challenge include the multi-channel front-end enhancement and diarization, training data augmentation, speech recognition modeling with multi-channel branches. Tested on the offical Eval$_1$ and Eval$_2$ set, our best system achieves a relative 34.3\% improvement in CER and 56.5\% improvement in cpCER, compared to the offical baseline system.
\end{abstract}

\section{Track \uppercase\expandafter{\romannumeral1}: Automatic Speech Recognition}
\label{sec:track1}
\begin{figure}[t] 
\centering 
\includegraphics[width=\linewidth]{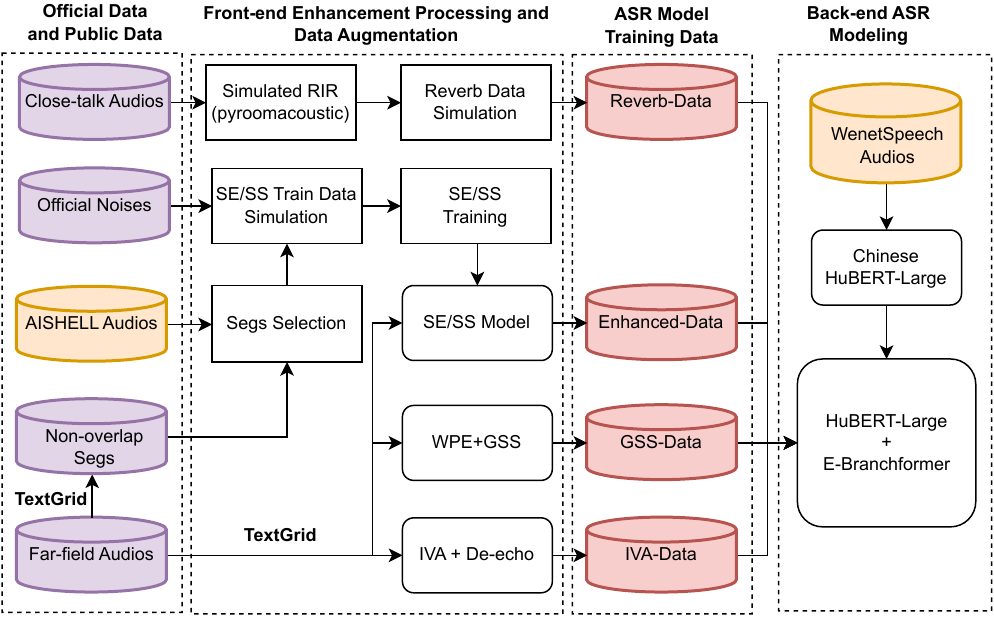} 
\vspace{-1em}
\caption{Illustration of our ASR training pipeline.}
\label{fig:training}
\vspace{-1em}
\end{figure}

Fig.~\ref{fig:training} illustrates our Automatic Speech Recognition (ASR) system training pipeline, including the utilization of several front-end enhancement modules, data simulation techniques, HuBERT representation model pre-training, and the final ASR model finetuning.

\subsection{Front-end Enhancement Processing}
\label{ssec:track1-frontend}
% Please revise.
To overcome the noisy recording environment of the speech data, we first implement several Speech Enhancement (SE) or Speech Separation (SS) models that include DCCRN~\cite{hu2020dccrn}, BSRNN~\cite{yu2022high}, GSS~\cite{raj2022gpu} and IVA~\cite{kim2006independent}. To train an enhancement model, we take non-overlap far-field training data and AISHELL1\footnote{AISHELL-1: https://www.openslr.org/33/} data as the clean data, which is then mixed with official noise data provided for noisy data simulation. As some far-field data contains loud in-car music, we filter out these far-field data with a pre-trained audio classification model~\cite{kong2020panns}. The training loss is SI-SDR. We trained both DCCRN and BSRNN enhancement models and found that BSRNN gave better results (higher SI-SDR). In addition, to separate the different speaker sources, we apply two statistic-based speech separation methods to solve the speaker overlap problem, including GSS and IVA methods. These enhanced and separated data from far-field recordings will be further used for back-end ASR data augmentation~(DA).
% for back-end ASR training. 
% Specifically, we select those Non-overlap segments from the far-field training data and combine them with AISHELL1~\cite{} dataset.  
% Then the segments are randomly selected to be merged and applied official noises to simulate data for SS model training. 
% \subsection{Data Augmentation}
% \label{ssec:track1-augment}
% To overcome the data sparsity problem, we utilize diverse data augmentation methods to enrich the training data. Specifically, except for the speech enhanced by the methods mentioned in section~\ref{ssec:track1-frontend}, 

Additionally, we employ Room Impulse Response~(RIR) to simulate the far-field speech from close-talk data with the toolkit named pyroomacoustic\footnote{Pyroomacoustic: https://github.com/LCAV/pyroomacoustics} where the room parameters are estimated by the given picture of the car. Further more, we add the real recorded noise into the RIR resulting speech to simulate more authentic far-field speech with the SNR between 0$\sim$10 dB. Finally, speed perturbation with the speed factor 0.9$\sim$1.1, and Spec-Augmentation are applied on-the-fly when training the ASR model. 

\subsection{Back-end ASR Training and Inference}
\label{ssec:track1-train}

Our ASR model is a joint architecture that constructs with a 24-layer HuBERT Encoder~\cite{HuBERT}, a 17-layer E-Branchformer Encoder and a 6-layer Transformer Decoder. 
The HuBERT-Large model was pre-trained by Tencent\footnote{HuBERT: https://github.com/TencentGameMate/chinese\_speech\_pretrain} with 10k hours Wenetspeech Chinese data. 

During decoding, GSS and IVA enhancement are first applied to the test far-field audios which then would be recognized with the ASR model. The final results are selected from either GSS or IVA enhanced parts based on the higher score of the ASR system for each sentence. Fig.~\ref{fig:test} (a) shows our inference pipeline for track \uppercase\expandafter{\romannumeral1}.

\section{Track \uppercase\expandafter{\romannumeral2}: Automatic Speech Diarization and Recognition}
\label{sec:track2}

In contrast to track \uppercase\expandafter{\romannumeral1}, track \uppercase\expandafter{\romannumeral2} would not provide any annotations, which indicates that the performance of Speaker Diarization~(SD) system is crucial to the final ASR results in this track. 

The speaker diarization system consists of two part: clustering-based method and modified TS-VAD. 
We follow the wespeaker~\cite{wang2023wespeaker} pipeline to achieve the spectral clustering based on the pretrained CAM++\footnote{CAM++: https://www.modelscope.cn/models/damo/speech\_campplus\_sv\_zh\-cn\_16k\-common/summary} for speaker embedding extraction, which is trained on around 200k Chinese speakers.
We then adopt modified TS-VAD framework. Specifically, we replaced the Bidirectional Long Short-Term Memory~(BLSTM) used in the original TS-VAD~\cite{medennikov2020target} with a four-layer transformer encoder. And the speaker embedding is extracted from pretrained CAM++~\cite{wang2023cam++} model for each speaker, instead of i-vector.
We utilize another CAM++ encoder to obtain the speech feature to replace the original MFCC feature.
For multi-channel input version, we use cross-attention module to leverage the information from four channel audios.

\begin{figure}[t] 
\centering 
\setlength{\abovecaptionskip}{0pt}
\setlength{\belowcaptionskip}{10pt}
\includegraphics[width=\linewidth]{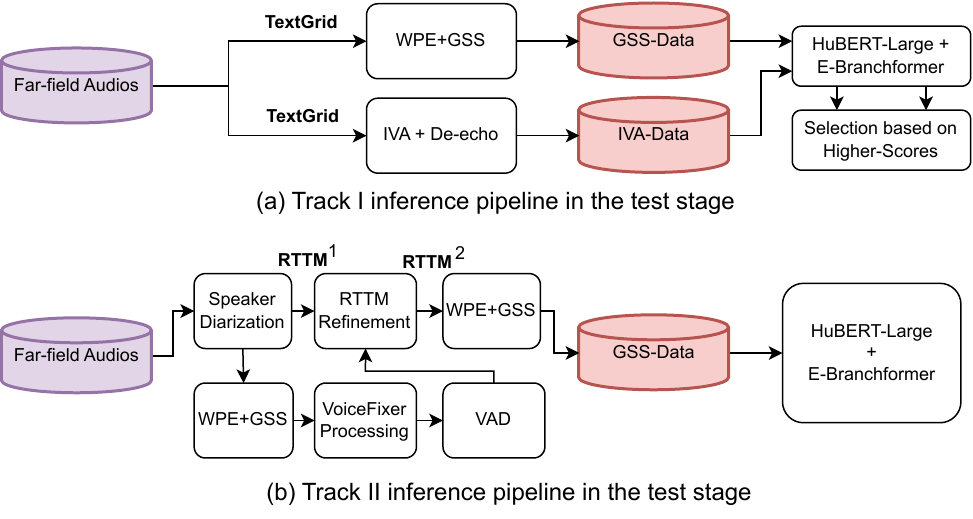} 
\vspace{-2em}
\caption{Illustration of inference pipelines in Track I and Track II.}
\label{fig:test}
\vspace{-1em}
\end{figure}

Considering that the our modified TS-VAD cannot perfectly segment far-field recordings, leading to numerous insertion errors in the final speech recognition. Motivated by this observation, we attempt to combine GSS and Voicefixer Toolkit\footnote{VoiceFixer: https://github.com/haoheliu/voicefixer} to further reduce the non-speech background noise. By employing VAD operation on the processed speech, we can distinguish the boundaries between speech and non-speech better, and update the original output RTTM$^1$ from the modified TS-VAD to the refined RTTM$^2$. Thus, we can run the GSS method again based on the refined RTTM$^2$ to get better speech for ASR. Fig.~\ref{fig:test} (b) shows the inference pipeline for track \uppercase\expandafter{\romannumeral2} while the ASR model is the exactly same ASR system in section~\ref{ssec:track1-train}.

\vspace{-1em}
\section{Experiment and Results}
\label{sec:exp}
% Brief introduction of our pipelines.

\subsection{Experiment Setup}
\label{ssec:exp-setup}

The 95 hours of Train set of ICMC-ASR challenge is utilized for jointly training the HuBert-E-Branchformer architecture ASR systems where the Dev set acts as the cross-validation set. The Eval$_1$ and Eval$_2$ sets are provided without transcription which is used for the final performance evaluation. The statistics information of these datasets is shown in Table~\ref{tab:ASR}. ESPnet\footnote{ESPnet: https://github.com/espnet/espnet} is used to train the joint ASR model. The learning rate is set to $5e^{-5}$ while the batchbins is 35M. The Warmuplr scheduler is applied with 40K warmup steps and the training target is hybrid CTC-Attention where the CTC weighting factor is set to 0.3. When perform decoding, the beamsize of beam-search is set to 10 and the CTC weight is 0.5. Average modeling is used and the best 5 models with lowest validation loss are selected.

\begin{table}[t]
    \centering
    \caption{Datasets for ASR model. WenetSpeech is for representation pretraining, and ICMC-ASR-Train is for ASR model training.}
    \begin{tabular}{c|c|c}
        \toprule
         Dataset & Duration (Hrs) & w/ Transcription  \\
        \midrule
        ICMC-ASR-Train & 95 & \ding{52} \\
        ICMC-ASR-Dev & 3.3 & \ding{52} \\ 
        ICMC-ASR-Eval$_1$ & 3.3 & \ding{55} \\
        ICMC-ASR-Eval$_2$ & 4.5 & \ding{55} \\
        \bottomrule
    \end{tabular}
    \label{tab:ASR}
    \vspace{-1em}
\end{table}

\subsection{Experimental Results}
\label{ssec:results}

The clustering-based method is used for original and rough diarization results. Based on that, we extract the non-overlap speech and speaker embedding for each speaker accordingly. These extracted speaker embeddings are fed into modified TS-VAD for further training. Observing the performance of V1 and V2 in Table~\ref{tab:dirization}, we conclude that taking 4-channel audios as input performs better due to more comprehensive information.  
V2 and V3 results demonstrate that second iteration can provide better diarization refinement.
We average the best three models from V3 experimental configuration and obtain V4 with the best performance.

Table~\ref{tab:asr-track1} illustrates the overall ASR results of our systems. A1 stands for the baseline system\footnote{Baseline System: https://github.com/MrSupW/ICMC\-ASR\_Baseline}, and A2 is the proposed joint architecture. For track \uppercase\expandafter{\romannumeral1}, our proposed ASR model achieves 32.6\% and 34.3\% performance improvement on Dev and Eval$_1$ sets respectively compared with the baseline system. And for track \uppercase\expandafter{\romannumeral2}, we choose V4 in Table~\ref{tab:dirization} as SD model, and A2 obtains 56.5\% cpCER performance improvement compared with A1.

\begin{table}[t]
    \centering
    \caption{Diarization results on Dev set.}
    \begin{tabular}{c|c|c|c|c|c|c}
        \toprule
         Index & Input & \# Iter. & MS & FA  & SC  & DER   \\
        \midrule
        V1 & DX02C01 & 1 & 4.38 & 5.16 & 4.25 & 13.79\\
        V2 & 4-Channel & 1 & 4.98 & 4.69 & 2.29 & 11.96\\
        V3 & 4-Channel & 2 & 4.87 & 2.34 & 1.28 & 8.49\\
        \textbf{V4} & 4-Channel & 2 & \textbf{3.00} & \textbf{2.89} & \textbf{0.26} & \textbf{6.15}\\
        
        \bottomrule
    \end{tabular}
    \label{tab:dirization}
    \vspace{-1.5em}
\end{table}

\begin{table}[t]
    \centering
    \caption{ASR results on Dev set, Eval$_1$ set and Eval$_2$ set.}
    \begin{tabular}{c|c|c|c|c|c}
        \toprule
        \multirow{2}{*}{Index} & \multirow{2}{*}{System} & \multicolumn{2}{c|}{Track I (CER)} & \multicolumn{2}{c}{Track II (cpCER)} \\ \cline{3-6}
          &   & Dev & Eval$_1$ & Dev & Eval$_2$ \\
        \midrule
        A1 & Baseline & 32.92 &26.24 & 65.90 &72.88\\
        % A2 & DA & 23.1 &17.68 & 34.30 &32.45\\
        A2 & Proposed & 22.2 &17.23 & - &31.68 \\
        \bottomrule
    \end{tabular}
    \label{tab:asr-track1}
    \vspace{-1em}
\end{table}

% Shown our Challenge results and diarization results?

% \begin{table}[htbp]
%     \centering
%     \begin{tabular}{c|c|c|c}
%         \toprule
%          Index & Sys. & Dev & Eval$_2$ \\
%         \midrule
%         A1 & Base & 65.90 &72.88 \\
%         A7 & DA & 34.30 &32.45 \\
%         A11 & Final & - &31.68 \\
%         \bottomrule
%     \end{tabular}
%     \caption{ASR results (cpCER) on Dev set and Eval$_2$ set.}
%     \label{tab:asr-track1}
% \end{table}

% \section{Conclusion}
% \label{sec:conclusion}
% In this paper, we present our system for the ICMC-ASR Challenge. THe 

% \section{acknowledgements}
% \label{sec:ack}

\vspace{-0.8em}
\bibliographystyle{IEEEbib}
\bibliography{refs}

\end{document}